\documentstyle[preprint,tighten,aps]{revtex}
\begin{document}           %
\draft
\preprint{\vbox{\noindent
Submitted to Astroparticle Physics\hfill astro-ph/9602045\\
          \null\hfill  INFNFE-01-96\\
          \null\hfill  INFNCA-TH9601}}
\title{Physics potentials of 
\\ pp and pep solar neutrino fluxes
      }
\author{
         E.~Calabresu,$^{(1)}$
         G.~Fiorentini$,^{(2)}$
         and M.~Lissia$^{(1)}$
       }
\address{
$^{(1)}$Istituto Nazionale di Fisica Nucleare, via Ada Negri 18,
        I-09127 Cagliari, Italy;\\ and
        Dipartimento di Fisica dell'Universit\`a di Cagliari, I-09124
        Cagliari, Italy\\
$^{(2)}$Dipartimento di Fisica dell'Universit\`a di Ferrara,
        I-44100 Ferrara, Italy;\\ and Istituto Nazionale di Fisica Nucleare,
        Sezione di Ferrara, I-44100 Ferrara, Italy
        }
\date{January 1996}
%
\maketitle                 
\begin{abstract}
Experimental determinations of the $pp$ and $pep$ fluxes have
great potentialities. We briefly review the reasons that make
such measurements privileged tests of neutrino properties.
We discuss the predictions for these fluxes given by four {\em good} 
solutions to the solar neutrino problem: small- and large-angle MSW and 
Just-So oscillations into active neutrinos, and small-angle
MSW oscillations into sterile neutrinos. In addition, we examine
the impact of the planned Hellaz detector, which should measure
separately the $\nu_e$ and $\nu_{\mu}$ fluxes in the $pp$ energy window and
the signal from the $pep$ neutrinos, for distinguishing among the different
solutions and for determining the solar central temperature.
\end{abstract}
\narrowtext
%
\subsection{Introduction}
\label{intro}
    Theoretical understanding of stellar structure has reached
a rather mature stage and can be confronted with refined experimental
tests in a wide range of conditions. In particular, present solar models
accurately reproduce even the very detailed experimental information
coming from helioseismology. Therefore, we are nowadays quite confident
in the predictions of standard solar models (SSMs) for the main neutrino 
fluxes, especially $pp$, $pep$ and $^7$Be 
neutrinos~\cite{BU88,Bahcall89,BP92,TC93,BP95}. 
Nevertheless, all the present experimental determinations of solar
neutrino fluxes are at odds with the theoretical predictions, strongly 
suggesting that neutrinos might have nonstandard 
properties~\cite{Bere93a,Bere95,TAUP95}.

    In this context, direct measurements of the $pp$ and $pep$ neutrino 
fluxes would be of the outmost importance. Predictions for these fluxes are 
the most robust and the least dependent on the detail of the solar models 
so that comparison with experiment could provide a powerful test of the 
different nonstandard-neutrino solutions. Moreover, future experiments, 
e.g. Hellaz~\cite{HELLAZ}, aimed to measure both the $\nu_e$ and 
$\nu_{\mu}$ flux would further increase the relevance of such a test. 
The significance of a measurement of the other main solar neutrino flux 
($^7$Be) has been discussed previously~\cite{TAUP95,Scilla,CFFL95}.

    Therefore, this paper aims to three main objectives:
\begin{itemize}
\item%
   to summarize what is known about the production of $pp$ and
   $pep$ neutrinos in the Sun;
\item%
   to examine predictions of different particle physics solutions to the 
   solar neutrino problem (SNP) for the fluxes and spectra of $pp$ and 
   $pep$ neutrinos at the Earth surface;
\item%
   to discuss the physics potential of a planned detector of $pp$ and
   $pep$ neutrinos (Hellaz) for distinguishing among the possible solutions  
   and for measuring the central temperature of the Sun.
\end{itemize}

\subsection{pp and pep neutrinos from the Sun}

   The $pp$-neutrino production rate from the Sun, $L_{pp}$, 
is predicted quite accurately; for example the 
last SSM of Bahcall and Pinsonneault~\cite{BP95} (BP95) yields:
\begin{equation}
    L_{pp}^{SSM} = 1.66\cdot(1\pm 0.01)\cdot 10^{38}\,\mbox{s}^{-1} \, .
\end{equation}
If neutrino are standard, i.e. electron neutrinos do not decay nor are 
converted to other flavors, this production rate determines the flux 
of $pp$ neutrinos on Earth at the distance $R_{TS}$ of one astronomical 
unit:
\begin{equation}
      \Phi_{pp}^{SSM}= 5.91\cdot(1\pm 0.01)\cdot 
                        10^{10}\,\mbox{cm}^{-2} \mbox{s}^{-1} \, .
\end{equation}
There is no surprise for such a small uncertainty, since 
the $pp$-neutrino production is strongly correlated with solar energy 
production, which is fixed by the presently observed luminosity. 
Actually, very simple considerations set an extremely reliable upper 
bound to $L_{pp}$, which turns out to be close to the SSM 
estimate. The maximum production rate $L_{pp}^{max}$ clearly
corresponds to the case that only $pp$ neutrinos are emitted from the Sun.
Given the $pp$-neutrinos average energy
$\langle E\rangle_{pp}=0.265$~MeV, the total energy released per
fusion $Q=26.73$~MeV and the observed solar luminosity~\cite{BP95}
$L_{\odot}= 2.399\cdot(1\pm 0.0042)\cdot 10^{39}\,\mbox{MeV s}^{-1}$, 
one finds:
\begin{equation}
L_{pp}^{max}  =  L_{\odot} / (Q/2 - \langle E\rangle_{pp})
              = 1.831 \cdot (1\pm 0.004)\cdot 10^{38}\,\mbox{s}^{-1}\, .
\end{equation}
Correspondingly, the maximal $pp$ flux on Earth is:
\begin{equation}
      \Phi_{pp}^{max}= 6.512\cdot (1 \pm 0.004)\cdot 
                        10^{10}\,\mbox{cm}^{-2} \mbox{s}^{-1} \, .
\end{equation}
The reason that $\Phi_{pp}^{SSM}$ is so close to this number is that 
SSM predicts that more that 90\% of the total number of neutrinos
are $pp$ neutrinos.

   Nonstandard solar models differ from the SSM and among themselves
because they have different physical inputs. Most of the effect of
changing some of these inputs is reasonably well parameterized by a single 
parameter: the central temperature $T$~\cite{Hata94,Castel94}.
For the sake of simplicity and concreteness, we shall often talk of 
changing the (central) temperature $T$ without referring to the specific 
way this change is obtained.

   When the temperature changes, the relative efficiencies of the different
chains ($pp$-I, $pp$-II, $pp$-III and CNO) also change, but the total neutrino 
production rate remains practically constant, since its value is strongly
constrained by the luminosity. As an example, if $T$ increases with respect 
to $T^{SSM}$, the $pp$-II chain becomes more efficient and yields a 
larger production of $^7$Be neutrinos; the CNO efficiency and neutrinos also
increase. Correspondingly, the $pp$ production decreases, even if its
dependence on the temperature is rather weak. By using a parameterization
of the form:
\begin{equation}
   L_{pp} = L_{pp}^{SSM}\,\left(\frac{T^{SSM}}{T}
                               \right)^{\beta_{pp}} \, ,
\end{equation}
the parameter $\beta_{pp}$ ranges from 0.6 to 0.85 depending on how the
temperature change is achieved~\cite{Castel94,Bahcall96}.

   This dependence implies that even a 5\% variation of $T$, which 
is a really huge variation on the scale of the SSM uncertainties
($\Delta T \approx 1\%$), changes $L_{pp}$ by just a few percent.

   The $pep$ neutrinos are estimated to be a tiny fraction of $pp$ neutrinos. 
Their ratio
\begin{equation}
\xi \equiv L_{pep}/L_{pp}
\end{equation}
is just $\xi = 2.37\cdot 10^{-3}$ in the SSM of BP.
   The corresponding flux on Earth, if all neutrinos survive, is
$
 \Phi_{pep}^{SSM}= 1.40\cdot 10^{8}\,\mbox{cm}^{-2}\mbox{s}^{-1} \, 
$.

 The value of $\xi$ is rather stable among the different SSM calculations 
(all SSMs give the same value of $\xi$ within about 10\%) and it is also 
weakly sensitive to the central solar temperature. By writing
\begin{equation}
   \frac{\xi}{\xi^{SSM}} = \left(\frac{T}{T^{SSM}}\right)^{\beta_{\xi}}\, ,
\end{equation}
the parameter $\beta_{\xi}$ takes values in the range from $-1.6$ to $2.8$, 
depending on the parameter which is varied to tune the central 
temperature~\cite{Castel94,Bahcall96}.

  The uncertainty on $\xi$ is at most 15\% for a 5\% variation of 
temperature, and again this should be taken as an extreme possibility 
for SSMs. 

   Concerning the energy distribution, kinematics fixes the energy of 
monochromatic $pep$ neutrinos ($E_{pep}=1.442$ MeV) and nuclear physics 
determines the shape of the $pp$ spectrum, essentially through phase 
space considerations; this latter spectrum is shown in Fig.~\ref{fig1} 
(solid curve).

   We remind that SSMs give also a very robust (stable) prediction for the 
flux of $^7$Be neutrinos. Nevertheless, the accuracy quoted for this flux 
is lower than the one for the $pp$ and $pep$ fluxes:
\begin{equation}
   \Phi_{Be}^{SSM} = 5.15 \cdot (1\pm 0.06) \cdot 
                            10^{9}\, \mbox{cm}^{-2}\mbox{s}^{-1}\, .
\end{equation}
This fact is also reflected in its somewhat stronger temperature dependence:
\begin{equation}
   \frac{\Phi_{Be}}{\Phi_{Be}^{SSM}} = 
                         \left( \frac{T}{T^{SSM}}\right)^{\beta_{Be}}\, ,
\end{equation}
where now $\beta_{Be}$ goes from about 8.7 to about 11.5.

   All in all, theoretical predictions for the number of $pp$ and $pep$ 
neutrinos emitted per second from the Sun look quite reliable and 
stable, and require really a minimum of solar physics, essentially energy 
conservation.
\subsection{The pp and pep neutrinos on Earth}
   Among the several particle physics solutions to the SNP,
mechanisms where $\nu_e$ oscillate into neutrinos of
other flavors (for the sake of definiteness $\nu_{\mu}$) or into sterile
neutrinos ($\nu_s$) are particularly appealing, as they require very little
adjustment of the minimal electro-weak standard model. 

In particular, matter enhanced (MSW) oscillations and ``Just-So'' 
oscillations give a simple and satisfactory description of the 
data~\cite{Krauss88,Phil92,Petc,hata,rossi,CFFL95}.

  In this context the available 
experimental results support only four specific solutions~\cite{TAUP95}, 
which are reported in Table~\ref{tbl1}.
Oscillations into active neutrinos provide good fits to the data
in the mass range relevant to the MSW 
mechanism both at small and large mixing-angle, and also in the mass
range relevant to Just-So oscillations. Oscillations into sterile neutrinos
give instead a good fit to the data only within the MSW small-mixing-angle
solution.

   The predicted fluxes of $\nu_e$ and $\nu_{\mu}$ in the $pep$ and $pp$ energy
regions (these latter integrated over their energy spectrum) at 
the best fit points are also shown in Table~\ref{tbl1}, whereas 
Figs.~\ref{fig2} and \ref{fig3} show the ranges of
$\Phi^{\mu}$ predicted at the 90\% CL by the four acceptable models.

   Concerning $pp$ neutrinos (see Fig.~\ref{fig2}), the MSW 
small-mixing-angle solution predicts a very low (strictly null if
neutrinos are sterile) $\nu_{\mu}$ signal: the $\nu_e$ flux is little or 
not at all suppressed both in the case of active and sterile neutrinos. 
In fact, it is well-known that MSW small-mixing-angle solutions of the SNP 
are characterized by a strong suppression of the $\nu_e$ flux
only in a small energy window centered at intermediate energies. Therefore, 
the study of solar neutrinos in the $pp$ energy range cannot distinguish
these solutions from the case of standard neutrinos.
On the other hand, the Just-So and MSW large-mixing-angle solutions predict 
a $\nu_{\mu}$ ($\nu_e$) flux on Earth about 40\% (60\%) of the SSM estimate.

   The situation looks significantly different for $pep$ neutrinos (see
Fig.~\ref{fig3}). The MSW small-angle cases predict a $pep$ $\nu_e$ 
flux that is essentially vanishing, since the $pep$ energy falls within the 
suppression window. Therefore, this solution gives a $\nu_{\mu}$ flux 
that is about equal to the $pep$ SSM flux, if neutrinos are
active. The MSW large-angle solution gives again a $\nu_{\mu}$ flux about
one half of the SSM one. On the other hand, the Just-So model can 
accommodate almost any value of the $pep$ $\nu_{\mu}$ ($\nu_{e}$) flux. 
In this respect, it is worth observing that a $\nu_{\mu}$ flux close
to zero (unsuppressed $\nu_{e}$ flux) is only acceptable in the context 
of Just-So oscillations or sterile neutrinos.

   Thus a simultaneous measurement of the $\nu_{\mu}$ (or $\nu_e$) flux 
for both $pep$ and $pp$ neutrinos has a remarkable discriminating power
among the various solutions, as it is clearly shown in Fig.~\ref{fig4}
where we present the expected ($\Phi^{\mu}_{pp}$,$\Phi^{\mu}_{pep}$), 
normalized to the SSM predictions. In particular, note that the MSW 
small-angle solution for active neutrinos, which cannot be 
distinguished from standard neutrinos when looking just at the $pp$ 
energy range, yields
drastically different predictions for the $pep$ neutrinos. In addition,
large parts of the 90\% CL regions predicted by MSW large-angle 
and Just-So oscillations uniquely characterize one of the two solutions, 
even if there is some ambiguity when $\Phi^{\mu}_{pp}$ is about 40\%
of $\Phi^{SSM}_{pp}$ and $\Phi^{\mu}_{pep}$ between 50\% and 80\%
of $\Phi^{SSM}_{pep}$. However, the distinguishing of sterile from
standard neutrinos needs independent information on the $\nu_{e}$ flux.

    A more complete picture of the possible outcomes 
and implications of an experiment capable of measuring both $\Phi_{e}$ 
and $\Phi_{\mu}$ is shown in Figs.~\ref{fig5} and \ref{fig6}, which we
are going to discuss in detail, for the $pp$ and $pep$ neutrinos, respectively.

    If we assume that $\nu_e$ transform into $\nu_{\mu}$ and/or 
$\nu_{s}$, the fluxes on Earth satisfy:
\begin{equation}
  \Phi^{e}_{pp} + \Phi^{\mu}_{pp} + \Phi^{s}_{pp} = 
     \frac{L_{pp}}{4\pi\, R_{TS}^2} \, .
\end{equation} 
We have already seen in the previous section that solar energetics provides 
an upper bound to $L_{pp}$ and, consequently, there exists an upper 
bound on the sum of electron and muon neutrinos:
\begin{equation}
      \Phi^{e}_{pp} + \Phi^{\mu}_{pp} 
           \leq  \Phi^{max}_{pp} \, ;
\label{lumlim}
\end{equation}
this upper bound is shown as a dashed line in Fig.~\ref{fig5}.
Thus, whatever be the mechanism responsible for the solar neutrino 
problem and independently of the SSM, experimental findings should stay 
below this line.

If one relies on the SSM prediction for the $pp$ flux $\Phi_{pp}^{SSM}$, 
one has instead:
\begin{equation}
     \Phi^{e}_{pp} + \Phi^{\mu}_{pp} \leq \Phi_{pp}^{SSM} \, .
\label{ssmlim}
\end{equation}
In Fig.~\ref{fig5}, this SSM bound is represented by the shaded band;
the width of this band indicates the uncertainty of the SSM $pp$ flux. 
This figure shows clearly 
that the SSM bound is not much lower than the more general bound 
coming from the luminosity constraint, Eq.~(\ref{lumlim}), in accordance 
with the well-known fact that most of total energy of SSM comes from the 
$pp$-I chain.

   The equal sign in Eqs.~(\ref{lumlim}) and (\ref{ssmlim}) only holds 
in the absence of sterile neutrinos, so that a measurement yielding
$\Phi^{e}_{pp}$ and $\Phi^{\mu}_{pp}$ along this line would be a clear 
indication against sterile neutrinos.

   Should instead the experiment give a point below this line, this 
result would imply either conversion into sterile neutrinos or a solar 
central temperature significantly higher than the one estimated by the 
SSM (we recall that as temperature increases $\Phi_{pp}$ decreases).

  The predictions of the four candidate solutions (best points and 90\%
CL regions) are also shown in Fig.~\ref{fig5}. While the sterile
neutrino solution lays on the horizontal axis, all three active
neutrino solutions lay along the shaded band that represents the
SSM prediction. The dashed arrow indicates the limit of the 90\% intervals 
on the band or on the horizontal axis.

   Similar considerations can be applied to $pep$ neutrinos, see 
Fig.~\ref{fig6}. We remark that the study of $pep$ neutrinos can clearly
discriminate between oscillations into active and sterile neutrinos for
the MSW small-angle solution (see Fig.~\ref{fig6}), whereas the
corresponding predictions for the $pp$ neutrinos are not
so clearly separated (see Fig.~\ref{fig5}). In addition, the study of 
$pep$ neutrinos is able to discriminate between Just-So and MSW 
large-angle solution
for a significant portion of the possible outcomes (compare the
90\% intervals predicted by these two solutions in Fig.~\ref{fig5}
with the corresponding intervals in Fig.~\ref{fig6}).

   In principle, measurements of the $pp$ neutrino energy spectrum could
provide additional information and help discriminating the different
solutions. The $pp$ $\nu_e$ energy spectra predicted by the different 
schemes are shown in Fig.~\ref{fig1} together with the one given by the 
SSM. Deformations of the spectrum are tiny for the MSW solutions, and the
only effect is basically a change of the normalization of the flux.
On the contrary, Just-So oscillations predict in principle a strong energy
dependence for the yearly averaged signal in the $pp$ region (solid
oscillating line in Fig.~\ref{fig1}). The spectral deformation is even
more clear in Fig.~\ref{fig7}~(a) where we plot the ratio of the $\nu_e$ 
flux to the SSM prediction as function of energy. 
In practice, however, a low energy resolution can miss this strongly 
oscillating energy deformation. The histograms (b), (c) and (d) in 
Fig.~\ref{fig7} have been made with bins of size $\Delta E=$ (a) 10, (b) 
20 and (c) 50~keV to estimate the necessary energy resolution.
While a bin size of 20~keV is still capable to resolve the energy dependence, 
a bin size of 50~keV seems insufficient to this purpose.

    For the Just-So case, we have the additional possibility of seasonal 
modulations, since the oscillation length is comparable to the Sun-Earth
distance. Detection of this effect would be a distinctive indication
of the Just-So mechanism. 

    However, when looking at $pp$ $\nu_e$, a 50~keV energy bin is sufficient 
to average out the phase of the modulation (see Fig.~\ref{fig7}) and, 
therefore, completely suppresses these seasonal modulations, since it 
is the same phase $\phi\sim \Delta m^2 * R / E$ that controls spatial
($R$) and energy ($E$) oscillations. The possibility of seeing these 
modulations in the $pp$ energy spectrum would require high statistics in 
small energy bins (about 10~keV). But, if such resolution and statistics
were available, the energy dependence of the signal would give a much 
better indication of Just-So oscillation than the seasonal variation,
since the energy spectrum can cover several wave lengths
($\Delta \phi \gg \pi$), while the change of the Earth-Sun distance
corresponds to only a fraction of the wave length ($\Delta \phi 
\leq \pi/4$ at $E=300$~keV).

    In this respect, the monoenergetic $pep$ neutrinos are more interesting.
Semiannual modulations as large as $\pm 35\%$ of the average $\nu_e$ flux
are possible, as it is exemplified by the dashed curve in Fig.~\ref{fig8}. 
It is worth remarking that these modulations are large when the 
suppression is about 50\% (as natural since the derivative
respect to the oscillation wavelength, and then to the energy, is
maximal), which is the case when this additional information is the most
valuable, since the yearly averaged information cannot discriminate between 
Just-So and MSW large-angle solutions, see Figs.~\ref{fig2} and
\ref{fig3}. On the contrary, the seasonal variation is minimal when
the $pep$ signal is either maximal or minimal, as it happens at the best 
fit point, but in this case the yearly average signal is sufficient for
discriminating among the solutions.

\subsection{The potential of the Hellaz experiment}

    The proposed Hellaz detector~\cite{HELLAZ} aims at measuring both the 
$\nu_e$ and $\nu_{\mu}$ fluxes in the $pp$ energy region,
$\Phi^{e}_{pp}$ and $\Phi^{\mu}_{pp}$, by exploiting the different
angular dependence of the $\nu_e$ and $\nu_{\mu}$ scattering
cross section on electrons. 
    These two fluxes should be separately determined. The expected
statistics should allow measurements in at least four energy bins. The
energy bins reported in the present proposal~\cite{HELLAZ} are shown
in Table~\ref{tbl2}, where we also show the predicted events 
according to the SSM. Note, however, that the energy resolution of the 
single events is expected to be higher, about 9--24~keV at $E=300$~keV,
than the width of these bins.

    For higher energy ($pep$ and $^7$Be) neutrinos, the difference
between $\nu_e$ and $\nu_{\mu}$ cross sections becomes less pronounced
and flavor discrimination is not possible. Then the experiment
determines just the following combination of fluxes:
\begin{equation}
   \Phi^{H}_{pep,Be} = \Phi^{e}_{i} + \alpha_{i} \Phi^{\mu}_{i} 
                    \quad\quad (i=pep,\,Be)\, ,
\label{signal}
\end{equation}
where $\alpha$ is the ratio of neutral current (NC) to NC plus 
charged current (CC) cross sections at the $pep$ or $^7$Be neutrino 
energy (1.442~MeV and 0.861~MeV, respectively). 
Approximately, these ratios are:
\begin{eqnarray}
   \alpha_{pep} &=& 1/5\\
\label{sigpep}
   \alpha_{Be}  &=& 1/4 \, .
\label{sigBe}
\end{eqnarray}

\subsubsection{What can Hellaz tell us about neutrino oscillations?}

   Part of the physics potential of Hellaz is indicated in the last 
three rows of Table~\ref{tbl1}, where we present $\Phi^{e}_{pp}$, 
$\Phi^{\mu}_{pp}$ and the $pep$ signal $S_{pep}$ at the best fit points 
for each solution.

   Of course, the muon signal is a privileged indicator of neutrino 
oscillations; note, however, that --- as already remarked --- the 
MSW small-angle solutions, for either sterile or active 
neutrinos, predict small or even vanishing muon signals in the $pp$ energy
region. In other words, these solutions look very much the same as if 
neutrinos were standard from the point of view of $pp$ neutrinos.
In this case, a simultaneous measurement of the $pep$ signal is 
particularly important, since it should clearly discriminate MSW 
small-angle solutions from standard neutrinos.

    The capability of discriminating among the various solutions is 
thus best understood from Fig.~\ref{fig9}, where we present the
correlation among the $pep$ signal $S_{pep}$ and the muon flux in the 
$pp$ energy region $\Phi^{\mu}_{pp}$. This figure is essentially similar in 
spirit to Fig.~\ref{fig4}, but it involves quantities that should be
directly measured by Hellaz. This combined measurement should 
unambiguously discriminate the two MSW small-angle solutions from
the others (and from each other). If the actual solution is the
Just-So mechanism, this combined measurement can distinguish it from
the MSW large-angle solution only if the $pep$ signal is close to its
SSM prediction, as it happens at the best fit point, or close to
maximal suppression.

   We remark that the study of the yearly averaged and energy integrated
signals cannot essentially distinguish between the Just-So and MSW 
large-angle mechanisms, at least for part of the possible values of
parameters $\Delta m^2$ and $\theta$ within their 90\% CL regions,
those values that yield a $pep$ signal between 40\% and 70\% the
SSM signal. As previously noted, however, these ambiguous cases might be
discriminated by the detection of the semiannual modulations that are
foreseen for the signal of the monochromatic neutrino lines. 
The time dependence of the survival probability 
$\Phi^{e}_{pep}/\Phi^{SSM}_{pep}$ (solid line) and of the corresponding
signal $S_{pep}/S^{SSM}_{pep}$ (dashed line) expected at the point of 
maximal variation (dashed line) are shown in Fig.~\ref{fig8}, as a 
characteristic example.
We remind that the variation is maximal when the average signal is most
ambiguous, i.e. about a half of the SSM value, and in this case a few
hundred events should be sufficient for a $3\sigma$ evidence. We
verified that 1000 SSM events should allow a $3\sigma$ detection of 
seasonal variations predicted by parameters in almost the whole 90\% 
confidence region of the Just-So solution. This number of SSM
events could be reached in a little more than one year of operation
if HELLAZ is filled with CF$_4$ gas as it has been proposed specifically
to detect these higher-energy neutrinos~\cite{Ypsilantis}.
Only when $S_{pep}$ is close to its maximum or minimum this statistics 
is not sufficient.

   In addition, one can exploit the information coming from the energy
dependence of the signal (spectral deformation).

   As anticipated in the general discussion in the previous 
sections, we do not expect a large spectral deformation corresponding to
the three MSW candidate solutions. In Table~\ref{tbl2}, we show the
calculated $\nu_e$ flux, averaged in each energy bin and normalized to the
same quantity in the SSM, at the best fit points of each solution.
Variations among the different bins are just a few per cent, which
are comparable to the expected statistical fluctuations. 

   However, it is possible to find values of the parameters
($\Delta m^2$ and $\theta$) within the 90\% CL region of the MSW 
small-angle solution (not the values at the best fit as we see in
Table~\ref{tbl2}) such that there is some detectable suppression, even
at the level of 30\%, of $\Phi^{e}$ in the highest energy bin relative
to the others. The physical explanation is that the suppression window
(which is a function of $\Delta m^2 / E$) for sufficiently small
neutrino masses can reach the upper part of the $pp$ energy spectrum.
The detection of such a deformation would provide a specific signature 
of this type of solutions, while not seeing such deformation would
further reduce the 90\% CL region of the MSW small-angle solution.

 For the Just-So candidate solution a large spectral deformation is
expected. Energy resolution is crucial in this case (see Fig.~\ref{fig7}). 
More specifically, an energy resolution between 9 and 24 keV for single
events, which should be obtained by Hellaz, should be able to detect
such deformation given the expected statistics~\cite{Ypsilantis}.
On the contrary, most of the structure is washed out by 50~keV
energy bins.

  We can summarize this last point saying that Hellaz should be capable of 
performing accurate solar neutrino spectroscopy in the extremely important 
low energy range and that there are concrete possibilities that in this
energy range there could be a detectable spectral deformation.

\subsubsection{What can Hellaz teach us about the Sun?}

   Apart from the problem of neutrino oscillations, detection of $pp$ 
and $pep$ neutrinos could also be very interesting for studying 
properties of the solar interior, and in this respect we would like to 
add the following comments. 

   Ignoring for the moment the possibility of conversion into sterile 
neutrinos, a measurement of $\Phi^{e}_{pp}$ and $\Phi^{\mu}_{pp}$ provides 
a measurement of the solar central temperature $T$, since:
\begin{equation}
   \Phi^{e}_{pp} + \Phi^{\mu}_{pp} = 
     \Phi_{pp}^{SSM} \left( \frac{T^{SSM}}{T} \right)^{\beta_{pp}} \, .
\end{equation}
where $\beta_{pp}\approx 0.7$.

    Hellaz, which should be able to determine the total flux within 15\%, 
could measure the solar temperature with a 20\% uncertainty. This 
uncertainty is much larger than the estimated theoretical one. Nevertheless, 
direct temperature determination of the solar interior looks fascinating.

    In principle, the $^7$Be neutrinos could be more precise indicators 
of the solar temperature, since the corresponding $\beta$ coefficient is
higher:
\begin{equation}
8.7\leq\beta_{Be}\leq 11.5 \, .
\end{equation}
A determination of these fluxes with 10\% accuracy would give
the temperature at the 1\% level. In the present proposal 
of Hellaz~\cite{HELLAZ} separate determinations
of $\Phi^{e}_{pep}$ and $ \Phi^{\mu}_{pep} $ are not foreseen, but it
is nevertheless interesting to keep in mind such a possibility.

   These statements hold ignoring the possibility of sterile neutrinos. 
Note however that all our favorite solutions do not predict any sterile
neutrinos in the $pp$ energy range (at least if one excludes the Hellaz 
highest energy bin). More generally, if one includes the possibility of 
sterile neutrinos, the above-derived values of $T$ should be interpreted 
as upper values on the true solar central temperature.

\subsection{Conclusions}
     Theoretical predictions for the production rate of $pp$ and 
$pep$ neutrinos in the Sun require little knowledge of solar
physics, basically energy conservation, and, therefore, are quite 
reliable and stable.
     Expectations for the $\nu_{\mu}$ flux at $pp$ and $pep$ energies
are summarized in Figs.~\ref{fig2} and \ref{fig3}, respectively, for 
different solutions to the solar neutrino problem.
     We remark the following points:

     (1) The combined measurement of both pp and pep neutrinos is 
extremely discriminating among the various solutions, see 
Fig.~\ref{fig4} and \ref{fig9}.

     (2) Deformations of the $pp$ energy spectrum, particularly those
predicted by the Just-So mechanism (Fig.~\ref{fig7}), should be
observable in detectors with the energy resolution such as that
claimed by Hellaz (9--24 KeV)~\cite{HELLAZ,Ypsilantis}.

     (3) Seasonal variations of the $pep$ signal, as predicted by the 
Just-So mechanism (Fig.~\ref{fig8}), should also be detected by Hellaz,
once it is filled with CF$_4$ to increase the statistics of these
higher-energy neutrinos. 

\acknowledgments
It is our pleasure to thank T.~Ypsilantis for several fruitful and 
informative discussions and for a critical reading of the manuscript.

After the completion of this paper we received a preprint by
J.~N.~Bahcall and P.~I.~Krastev~\cite{BahKra} that contains material that 
partially overlaps with the one discussed in present work.
\begin{table}
\caption[taa]{
Predictions for $pep$ and pp neutrinos. For four different solutions
we present, at the best fit point 
($\Delta m^2$, $\sin^2{2\theta}$), the $\chi^2$ per degree of 
freedom, the fluxes of $\nu_{e}$ and 
$\nu_{\mu}$ originating from $pep$ and $pp$ reaction in units of SSM fluxes. 
For $pep$ neutrinos we also present the fraction of the SSM signal predicted 
for a CC+NC detector.
\label{tbl1}
               }
\begin{tabular}{lcccc}
&  \multicolumn{3}{c}{active}   &  sterile  \\
\cline{2-4}
&   MSW small $\theta$ & MSW large $\theta$ & Just So
                                  & MSW small $\theta$           \\
\tableline
$\Delta m^2$ [eV$^2$]& $7.9\cdot 10^{-6}$ & $1.7\cdot 10^{-5}$ 
                       & $6.0\cdot 10^{-11}$ & $4.9\cdot 10^{-6}$ \\
$\sin^2{2\theta}$ & $5.8\cdot 10^{-3}$ & 0.63 & 1.00 & $7.9\cdot 10^{-3}$ \\
$\chi^2/d.o.f.$~\tablenote{For comparison standard neutrinos yield
$\chi^2/d.o.f. = 904.1/4$.}
& 0.9/2 & 1.5/2 & 1.9/2 & 0.7/2 \\
\tableline
$\Phi^{e}_{pep}/\Phi^{SSM}_{pep}$ & 0.056 & 0.408 & 0.013 & 0.017 \\
$\Phi^{\mu}_{pep}/\Phi^{SSM}_{pep}$ & 0.944 & 0.592 & 0.987 & 0.983 \\
\tableline
$\Phi^{e}_{pp}/\Phi^{SSM}_{pp}$ & 0.986 & 0.643 & 0.508 & 0.979 \\
$\Phi^{\mu}_{pp}/\Phi^{SSM}_{pp}$ & 0.014 & 0.357 & 0.492 & 0.021 \\
\tableline
$S_{pep}/S^{SSM}_{pep}$ & 0.245 & 0.526 & 0.211 & 0.017 \\
\end{tabular}
\end{table}
\begin{table}
\caption[tbb]{
The first column shows the number of events expected in the Hellaz 
detector after one year of operation~\cite{HELLAZ} for standard neutrinos
and SSM. The next four columns show the $\nu_e$ survival probability
according to the four solutions of Table~\ref{tbl1}. The first
five rows report results for the indicated energy bins in the pp energy 
region and the last two rows for the $^7$Be and $pep$ lines.
\label{tbl2}
               }
\begin{tabular}{lccccc}
& number of events     &            
\multicolumn{4}{c}{$\Phi^{e}/\Phi^{SSM}$}   \\
\cline{3-6}
& for  &            \multicolumn{3}{c}{active}   &  sterile  \\
\cline{3-5}
& standard $\nu$  & MSW small $\theta$ & MSW large $\theta$ & Just So
                                  & MSW small $\theta$           \\
\tableline
energy bins [keV]& & & & & \\
 220--270 & 349 & 0.99 & 0.66 & 0.49 & 0.99 \\
 270--320 & 634 & 0.99 & 0.65 & 0.49 & 0.99 \\
 320--370 & 840 & 0.98 & 0.64 & 0.56 & 0.98 \\
 370--420 & 620 & 0.98 & 0.63 & 0.47 & 0.96 \\
 220--420 &2443 & 0.99 & 0.65 & 0.51 & 0.98 \\
lines ([MeV]) & & & & & \\
$^7$Be (0.861) & 1500 & 0.09 & 0.48 & 0.78 & 0.02 \\
$pep$ (1.442)     &  100~\tablenote{For these more energetic neutrinos,
it has been proposed to fill HELLAZ with CF$_4$, which would give about
850 events per year.}
& 0.06 & 0.41 & 0.01 & 0.02 \\
\end{tabular}
\end{table}
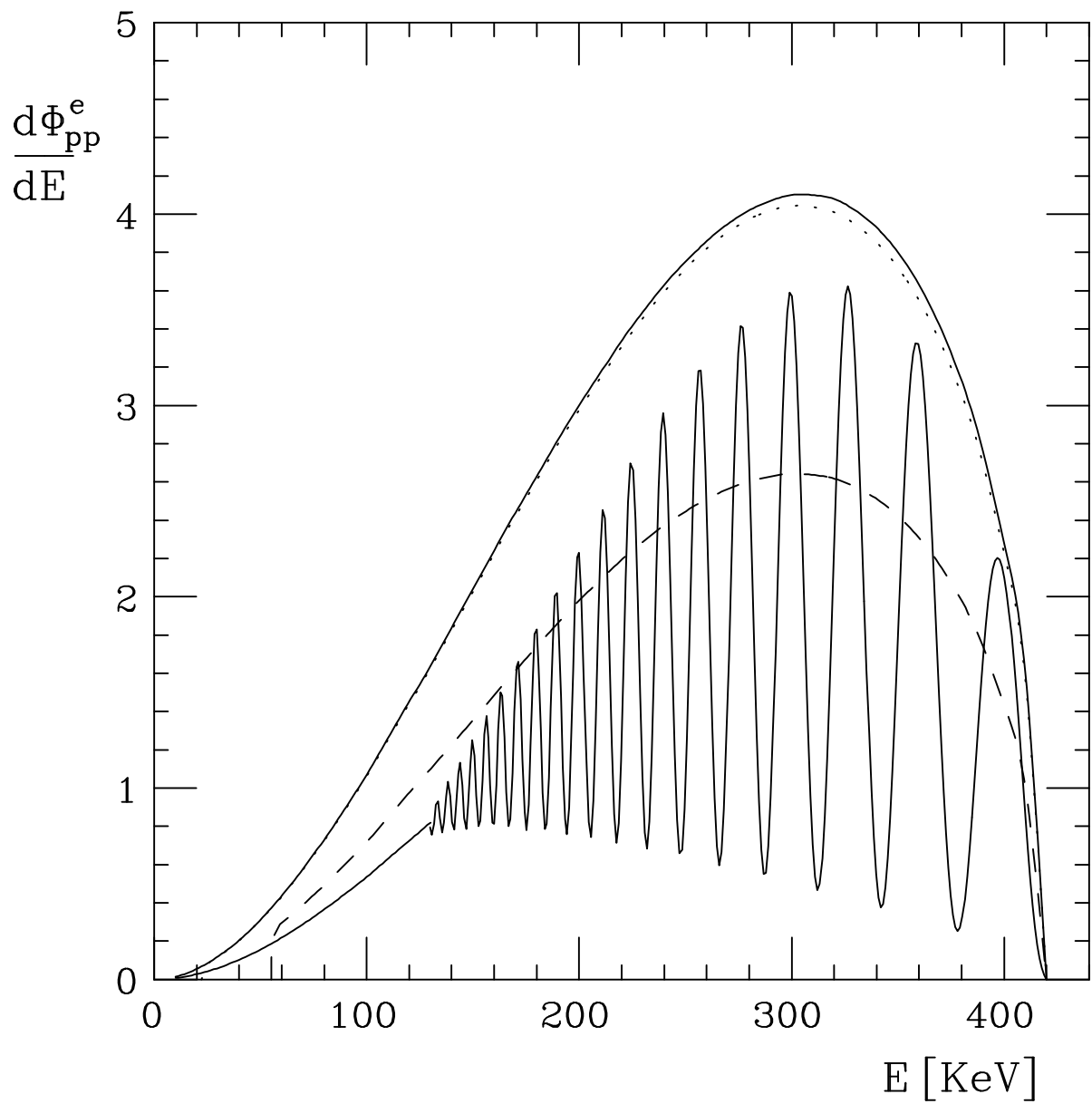
\begin{figure}
\caption[faa]{
        The yearly averaged $\nu_e$ spectrum on Earth in the $pp$ energy 
region
($0<E<440$~keV) for SSM (smooth solid line), MSW global best fit point 
in the small-$\theta$ region (dotted line), MSW local best fit point in the 
large-$\theta$ region (dashed line) and Just-So best fit (oscillating solid
line). The spectrum of the small-$\theta$ solution for sterile neutrinos 
is not distinguishable from the one for active neutrinos (dotted line).
For graphical reasons, the $\nu_e$ spectrum predicted by the Just-So best 
fit has been averaged below 150~keV eliminating the oscillation, which
has an amplitude (frequency) that decreases (increases) as the energy
decreases. The spectra are normalized such that the SSM spectrum 
integrates to one.
               }
\label{fig1}
\end{figure}
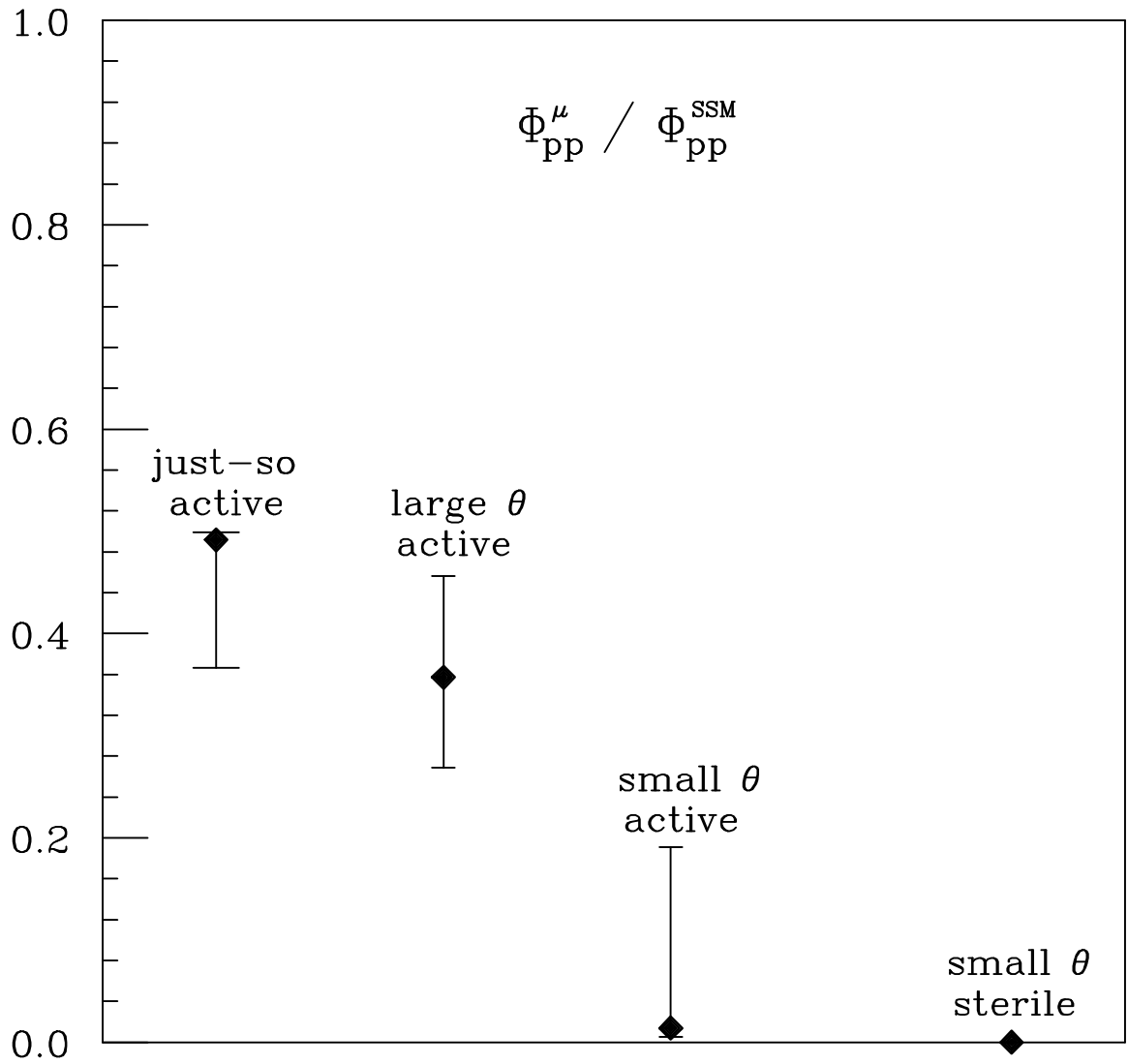
\begin{figure}
\caption[fbb]{
    The ratio $\Phi^{\mu}_{pp}/\Phi_{pp}^{SSM}$ predicted by the four 
{\em good} solutions in $pp$ energy range. Diamonds indicate the best 
fit predictions and error bars show the range of values allowed at the 
90\% CL.
               }
\label{fig2}
\end{figure}
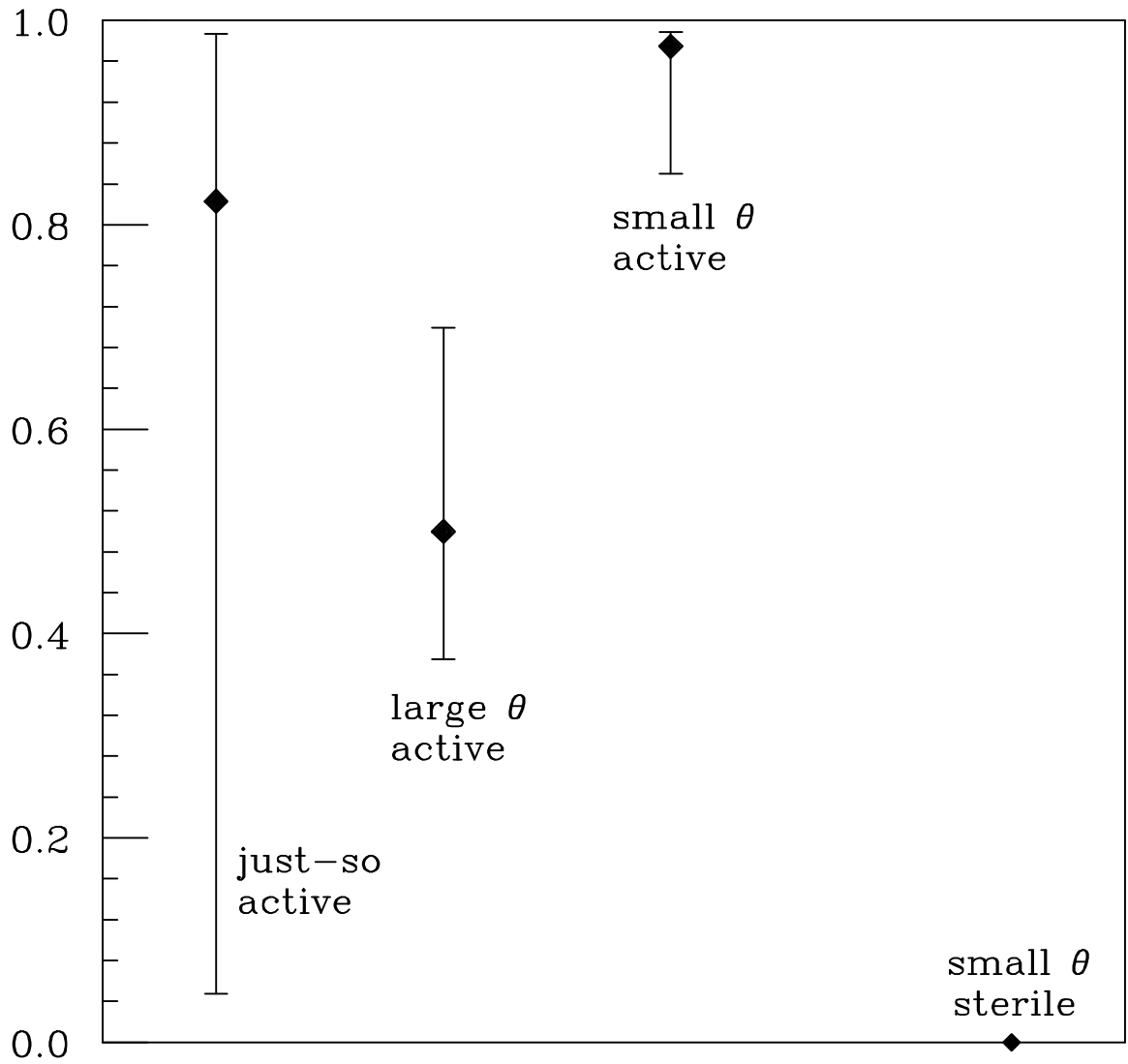
\begin{figure}
\caption[fcc]{
   Same as Fig.~\ref{fig2} for $pep$ neutrinos.
               }
\label{fig3}
\end{figure}
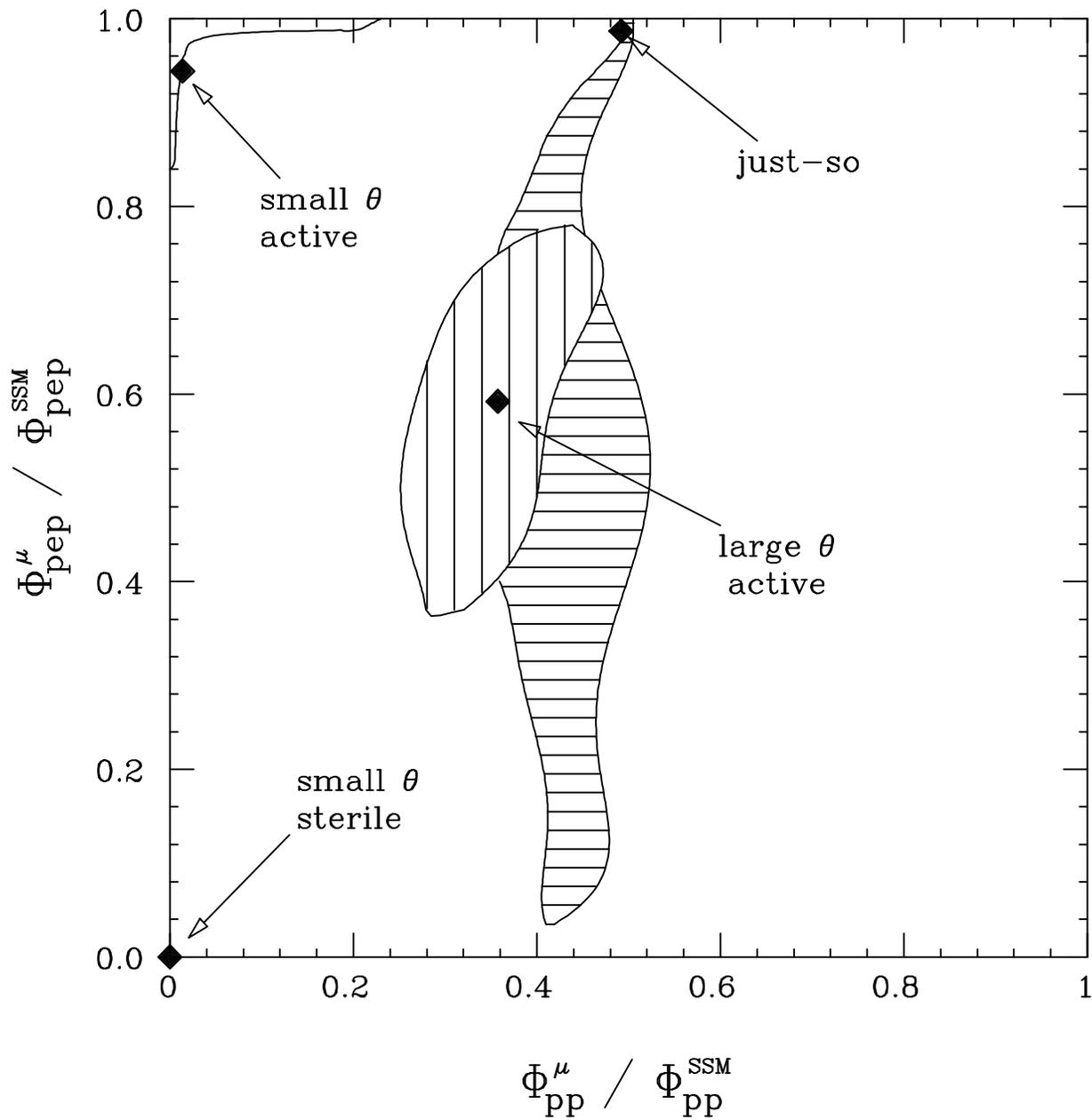
\begin{figure}
\caption[fdd]{
   The ratio $\Phi^{\mu}_{pep}/\Phi_{pep}^{SSM}$ vs. the ratio 
$\Phi^{\mu}_{pp}/\Phi_{pp}^{SSM}$ as they are predicted by the three 
{\em good} solutions to the SNP with oscillations into active neutrinos.
Diamonds indicate the predictions at the best fit parameters. Areas contain 
the corresponding ranges of parameters allowed at the 90\% CL. Oscillations 
into sterile neutrinos obviously give only the point at the origin.
               }
\label{fig4}
\end{figure}
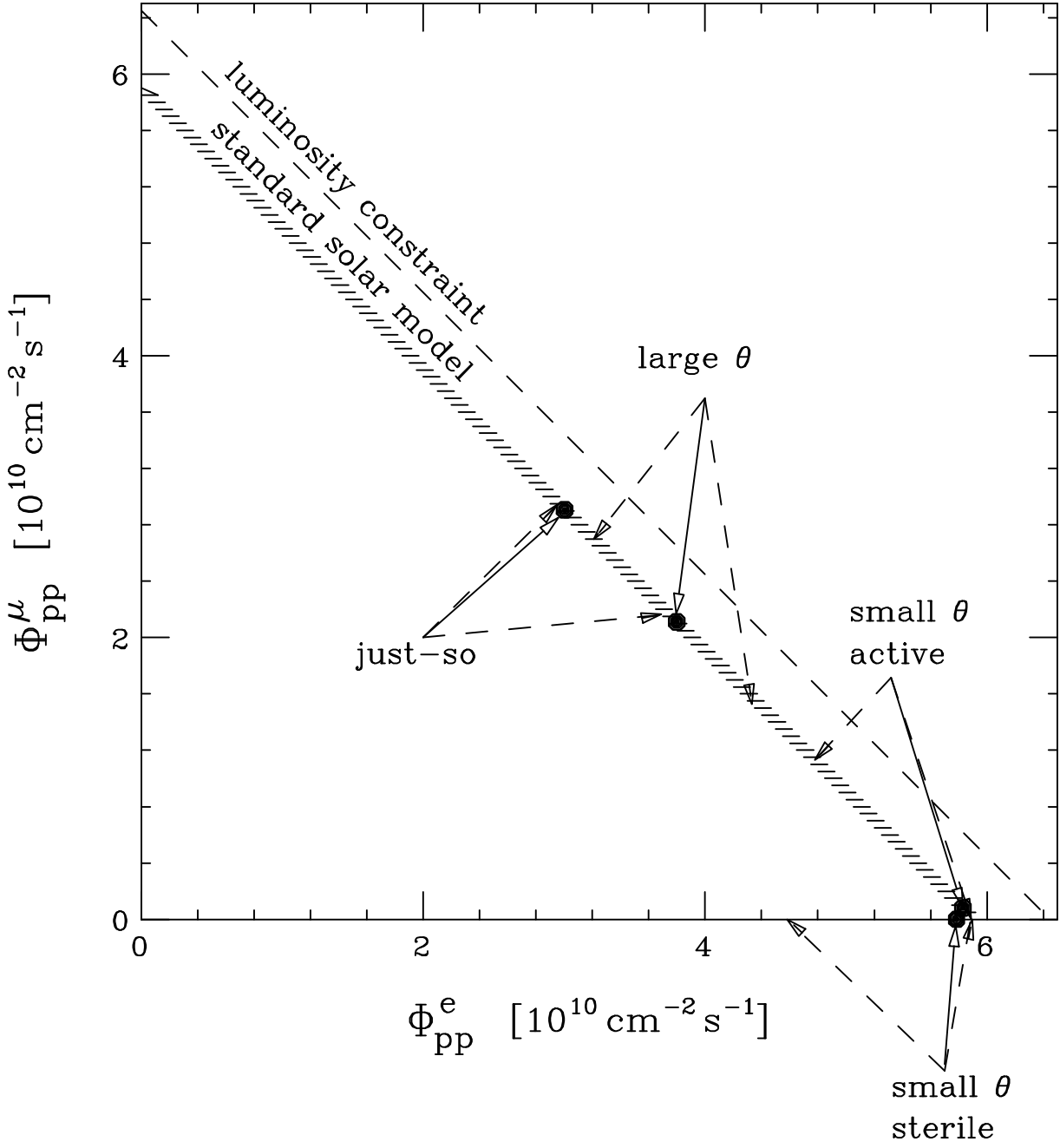
\begin{figure}
\caption[fee]{
   Possible outcomes of a combined measurement of $\Phi_{pp}^{e}$ and 
$\Phi_{pp}^{\mu}$. The dashed line is the upper bound due to the 
luminosity constraint, Eq.~(\ref{lumlim}). The shaded band shows the 
range of predictions allowed by the uncertainties of the SSMs in case of 
oscillations into active neutrinos. The solid arrows
indicate the prediction at the best fit parameters for the four solutions
considered. The dashed arrows 
show the corresponding ranges of parameters allowed at the 90\% CL. 
               }
\label{fig5}
\end{figure}
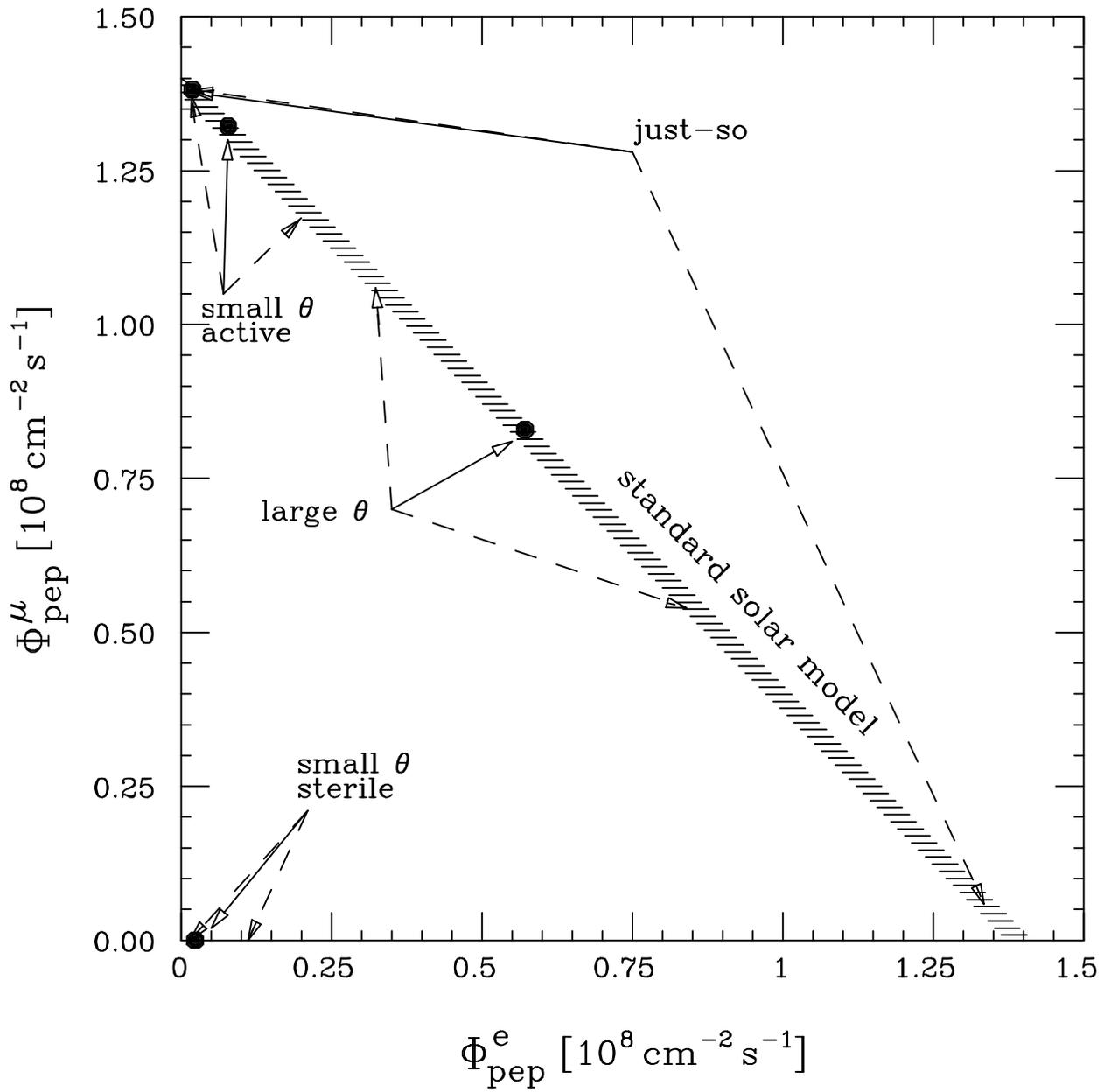
\begin{figure}
\caption[fff]{
   Same as Fig.~\ref{fig5} for $pep$ neutrinos.
               }
\label{fig6}
\end{figure}
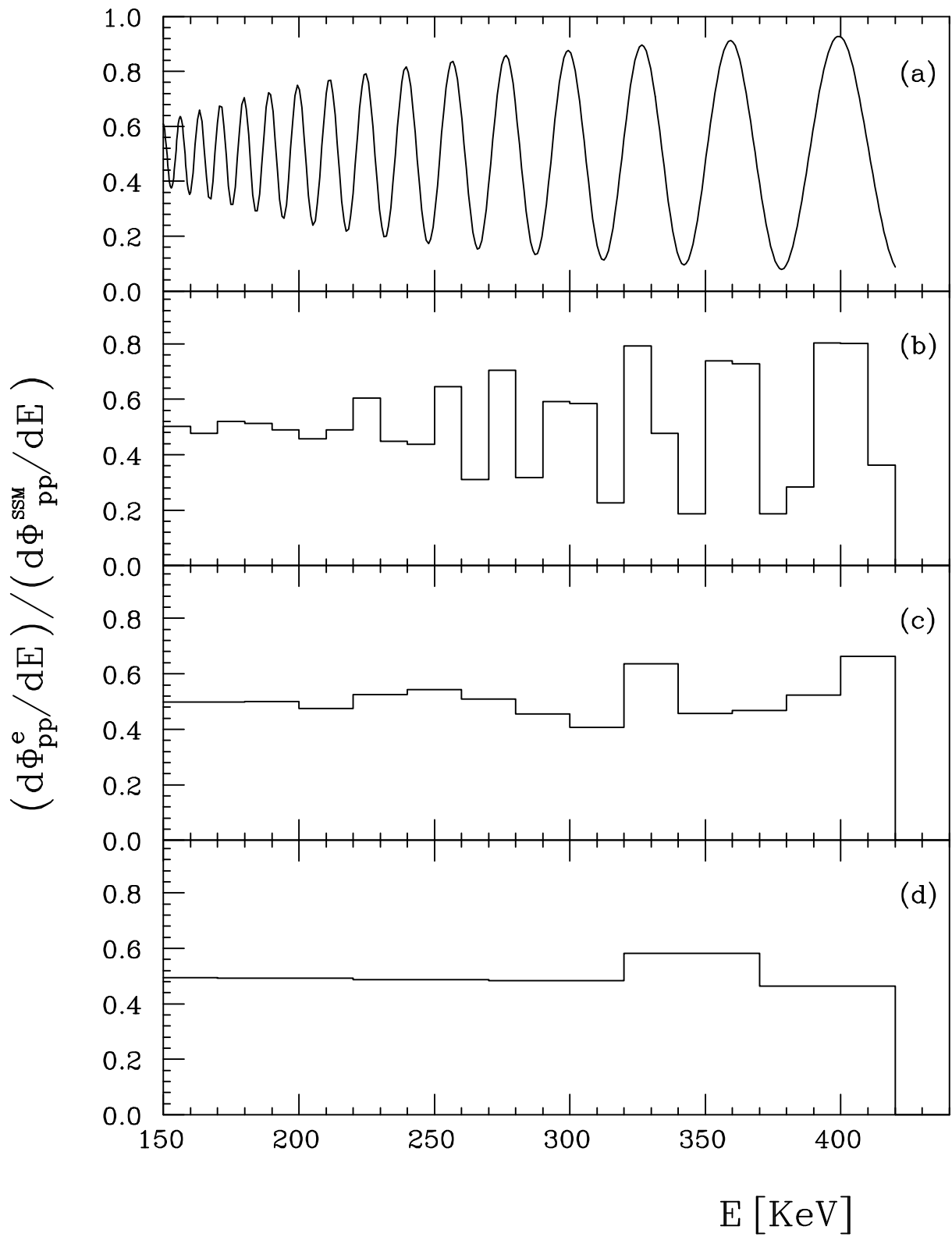
\begin{figure}
\caption[fgg]{
        The ratio of the $\nu_e$ spectrum on Earth in the $pp$ energy 
region ($150<E<440$~keV) predicted by the Just-So best fit (a)
for active neutrinos over the spectrum of the SSM. 
The effect of averaging the spectrum predicted by the Just-So best fit 
solution over an energy window of $\Delta E=$ (b) 10, (c) 20 and (d) 50~keV 
is also shown.
               }
\label{fig7}
\end{figure}
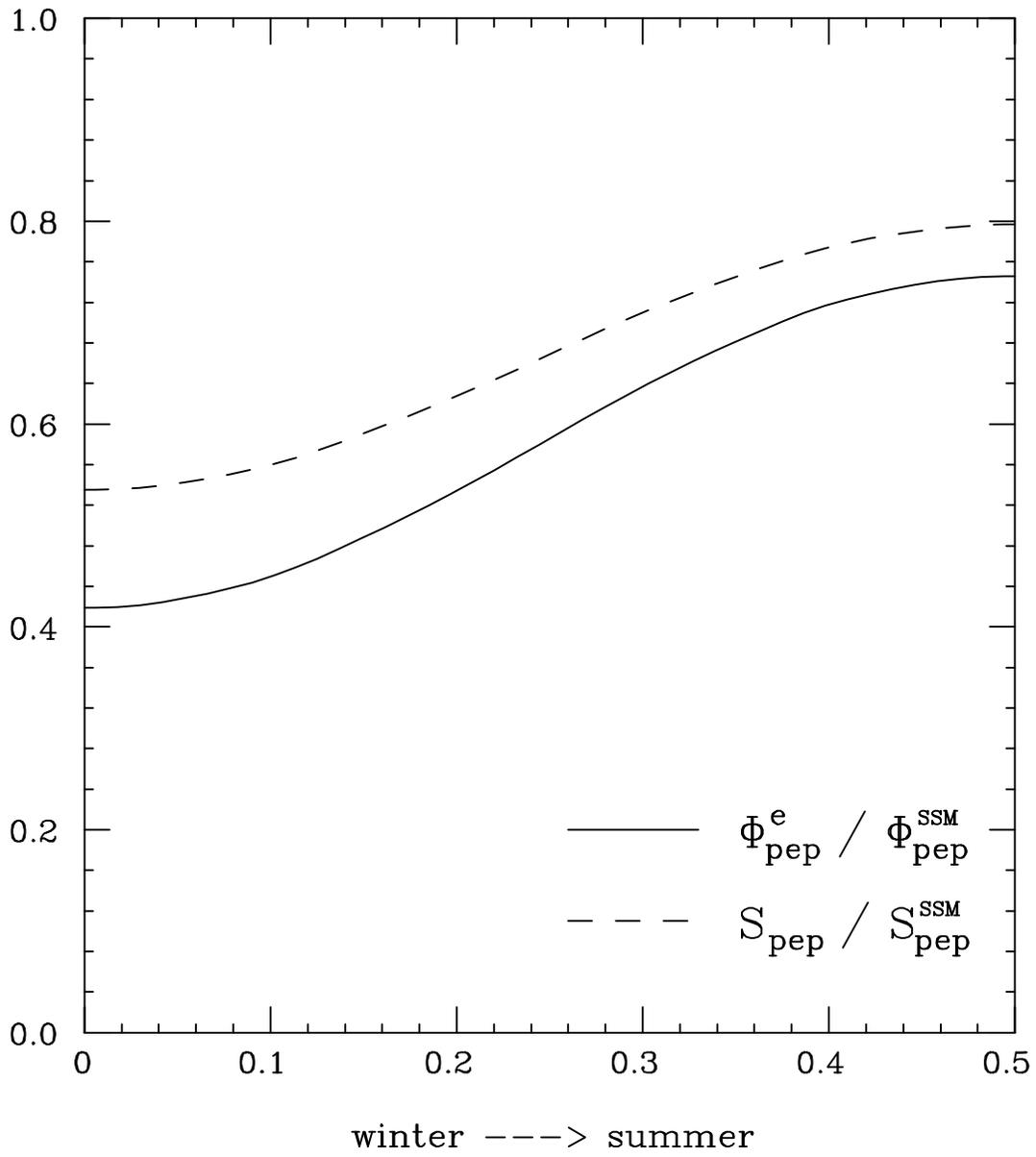
\begin{figure}
\caption[fhh]{
The expected time dependence of the $pep$ electron neutrino survival
probability $\Phi^{e}_{pep}/\Phi^{SSM}_{pep}$ (solid line)
as function of the absolute time difference from the perihelion in fraction 
of year (0.5 is then the aphelion). The dashed line shows instead
the signal $S_{pep}/S^{SSM}_{pep}$, see Eqs.~(\ref{signal}) and
(\ref{sigpep}). The oscillation parameters, $\Delta m^2$ and $\theta$,
are those that give the maximal variation within the 90\% confidence 
region.
               }
\label{fig8}
\end{figure}
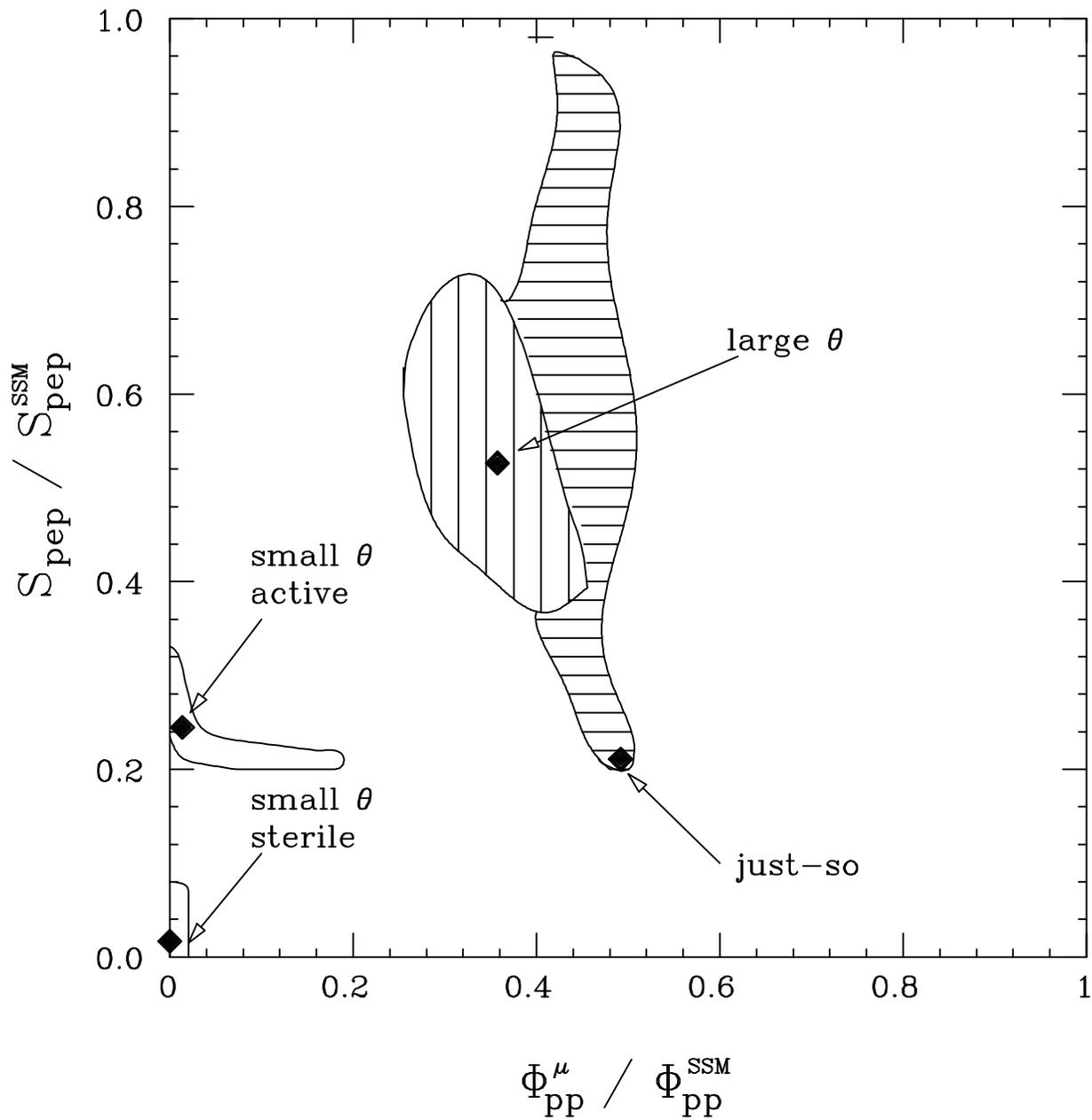
\begin{figure}
\caption[fii]{
Same as Fig.~\ref{fig4} with the ratio $\Phi^{\mu}_{pep}/\Phi_{pep}^{SSM}$
replaced by the quantity that should be measured by Hellaz
$S_{pep}/S^{SSM}_{pep}$. In this case, oscillations into sterile neutrinos 
give a non-zero value for $S_{pep}/S^{SSM}_{pep}$ and the range of values
allowed at the 90\% CL by this solution lies along the vertical axis.
               }
\label{fig9}
\end{figure}
\end{document}